\documentclass[prl,amsmath,amssymb,amsfonts,twocolumn,nofootinbib]{revtex4-1}
\usepackage{bm,graphicx,mathrsfs}
\usepackage{graphicx}
\usepackage{epsfig}
\usepackage{amsmath,bbm}
\usepackage{amsfonts,amssymb}
\usepackage{times}

\usepackage{color}
\usepackage{verbatim}

\newcommand{\assign}{:=}

\newcommand{\bR}{\mathbbm{R}}

\newcommand{\bZ}{\mathbbm{Z}}

\newcommand{\bE}{\mathbbm{E}}
\newcommand{\bL}{\mathbbm{L}}

\newcommand{\bC}{\mathbbm{C}}
\newcommand{\cL}{{\cal L}}
\newcommand{\cQ}{{\cal Q}}
\newcommand{\cO}{\mathcal{O}}
\newcommand{\cS}{{\cal S}}

\newcommand{\cM}{{\cal M}}
\newcommand{\cP}{{\cal P}}

\newcommand{\cE}{{\cal E}}
\newcommand{\cG}{{\cal G}}
\newcommand{\cD}{{\cal D}}

\newcommand{\1}{\mathbbm{1}}

\newcommand{\tr}[1]{{\rm tr}\left({#1}\right)}
\newcommand{\ptr}[2]{{\rm tr}_{#1}\left({#2}\right)}

\newcommand{\raw}{\rightarrow}

\newcommand{\Var}{{\mbox{Var}}}
\newcommand{\oket}[1]{ | \, #1  ) }
\newcommand{\obra}[1]{ ( #1 \,  |}
\newcommand{\oproj}[1]{ |#1) ( #1 |}
\newcommand{\ket}[1]{|#1\rangle}

\newcommand{\avr}[1]{\langle#1\rangle}
\newcommand{\half}{\frac{1}{2}}
\def\>{\rangle}
\def\<{\langle}

\newcommand{\Sp}{\,\,\,\,\,\,}
\newcommand{\no}{\nonumber\\}

\newcommand{\be}{\begin{equation}}
\newcommand{\ee}{\end{equation}}
\newcommand{\bq}{\begin{eqnarray}}
\newcommand{\eq}{\end{eqnarray}}
\newcommand{\bea}{\begin{eqnarray}}
\newcommand{\eea}{\end{eqnarray}}

\newtheorem{theorem}{Theorem}
\newtheorem{lemma}[theorem]{Lemma}

\newtheorem{proposition}[theorem]{Proposition}

    {\hspace*{\fill}$\Box$\vspace{1.5ex}\par}
    
\newcommand{\qed}{}

\def\Proof{\noindent{\it Proof.}}
\def\proof{\Proof}
\def\qed{\leavevmode\unskip\penalty9999 \hbox{}\nobreak\hfill
     \quad\hbox{\leavevmode  \hbox to.77778em{%
               \hfil\vrule   \vbox to.675em%
               {\hrule width.6em\vfil\hrule}\vrule\hfil}}
     \par\vskip3pt}
     
\begin{document}

\definecolor{james}{rgb}{1,.6,0}
\newcommand{\mjk}[1]{{\color{james} #1}}
\newcommand{\je}[1]{{\color{green} #1}}

\title{How fast do stabilizer Hamiltonians thermalize?}

\author{Kristan Temme$^1$ and Michael J. Kastoryano$^2$}
\affiliation{$^1$ Institute for Quantum Information and Matter, California Institute of Technology, Pasadena, CA 91125, USA\\
		$^2$ NBIA, Niels Bohr Institute, University of Copenhagen, 2100 Copenhagen, Denmark}
\date{\today}
\begin{abstract}
We present rigorous bounds on the thermalization time of the family of quantum mechanical spin systems known as stabilizer Hamiltonians. The thermalizing dynamics are modeled by a Davies master equation that arises from a weak local coupling of the system to a large thermal bath. Two temperature regimes are considered. First we clarify how in the low temperature regime, the thermalization time is governed by a generalization of the  energy barrier between orthogonal ground states. When no energy barrier is present the Hamiltonian thermalizes in a time that is at most quadratic in the system size.  Secondly, we show that above a universal critical temperature, every stabilizer Hamiltonian relaxes to its unique thermal state in a time which scales at most linearly in the size of the system. We provide an explicit lower bound on the critical temperature. Finally, we discuss the implications of these result for the problem of self-correcting quantum memories with stabilizer Hamiltonians. 
\end{abstract}
\maketitle

The study of thermalization times of Markov processes for classical spin systems is by now a mature research field, which has lead to important contributions for stochastic optimization problems, image reconstruction and a better understanding of Monte Carlo algorithms. Aside from these applications, the analysis is interesting in its own right as it has lead to a better understanding of non-equilibrium thermalization processes typically modeled by Glauber dynamics \cite{Martinelli}. In this letter, we study quantum mechanical Markov processes that converge to the Gibbs state of interacting quantum spin Hamiltonians. We are in particular interested in the equilibration time of a specific type of master equation, the Davies generator \cite{Davies}, which can be seen as a physically motivated quantum generalization of classical Glauber dynamics. We prove thermalization time bounds for the important class of stabilizer Hamiltonians. These models are well motivated by the study of \textit{self correcting quantum memories} (SCQM) and the bounds provide rigorous no go results for certain candidate models. We show that at low temperatures the convergence is determined by a generalized energy barrier (Eqn. (\ref{genEnBar})) of the stabilizer Hamiltonian -- a quantity that only depends on the excitations of the Hamiltonian. Furthermore, we give a bound on the critical temperature above which the thermal process always equilibrates rapidly and is independent of the low energy excitations. 

We bound the thermalization time of a quantum dynamical semigroup generatored by a \textit{Liouvillian} $\cL$ in terms of the \textit{mixing time}. If the steady state $\rho$ of $\cL$ is the Gibbs state of a Hamiltonain $H$, then the mixing time $t_{mix}$ is the least time such that $|| \varphi(t)-\rho ||_1\leq e^{-1}$, for any initial state $\varphi$ and $t \leq t_{mix}$, where $\varphi(t)=e^{t\cL^*}(\varphi)$, and the semigroup converges to the Gibbs state $\rho=e^{-\beta H}/\tr{e^{-\beta H}}$.  Here and in the following, $\cL^*$ refers to the dynamics in the Schr{\"o}dingier picture, and to $\cL$ in the Heisenberg picture (see Eqn. (\ref{DaviesGenerator})). It can be shown in general \cite{chi2}, that $ ||\varphi(t) -\rho||_1\leq \sqrt{||\rho^{-1}||}e^{-t \lambda} $, where $||\rho^{-1}||\leq e^{\beta ||H||}$ is the inverse of the smallest eigenvalue of $\rho$ and $\lambda$ is the spectral gap of the Liouvillian $\cL$, defined as the smallest (in magnitude) non-zero real part of an eigenvalue of $\cL$. Since $||H||$ scales as the volume of the system, the mixing time primarily depends on the gap of the Liouvillian: $t_{mix} \sim \cO( \lambda^{-1} \beta ||H||)$. Hence our main theorems will consist of lower bounding the spectral gap of the generators $\cL$ in different settings.

A central motivation for this work is the investigation of thermal stability of candidate Hamiltonians that can serve as a SCQM \cite{Dennis}. Many of the quantum spin Hamiltonians we consider are topologically ordered \cite{Kitaev} and can encode  quantum information in a degenerate subspace, for which the states can not be distinguished by local operations. Ideally, a SCQM should be able to store an arbitrary   quantum state in contact with a thermal bath that survives for a time which grows exponentially in the size of the system below some critical temperature. In other words, the code subspace would need to be metastable. So far, only one candidate system is known which satisfies this definition: the 4D toric code \cite{Dennis}. In two or three dimensions a number of no-go results exist \cite{NoGo1,NoGo2,NoGo3,NoGo4,TopOrderFiniteT}, which exclude the possibility of a quantum memory for certain classes of Hamiltonians. Most of these results rely on heuristic assumptions, such as the validity of the Arrhenius law \cite{QMemoriesReview}. The bounds on the thermalization time in this paper provide rigorous no go results, since $t_{mix}$ provides a natural upper bound to the lifetime. If the system thermalizes in a time that scales polynomially in the system size, we will say the system is not a good quantum memory. Note that this approach only permits to rule out possible candidates since a slow thermalization rate can only indicate the ability to store classical information, while quantum information may have already been lost.  


\textit{Preliminaries}: 
The stabilizer formalism was introduced by Gottesman \cite{Gottesman} as a convenient framework to express quantum error correcting codes (QECC).
It soon became clear that this formalism could also express a large family of commuting, frustration-free Hamiltonians, with the toric code as a prime non-trivial example \cite{Kitaev, Dennis}. 
\textit{The Pauli group} on $N$ qubits is defined as the set $\mathcal{P}_N =\{ e^{i\varphi} \sigma^{\alpha_1}_1\otimes \ldots \otimes \sigma^{\alpha_N}_N: \alpha_j\in\{I,X,Y,Z\}, \Sp \varphi \in \{ 0, \pi/2, \pi, 3\pi/2 \} \}$ with the group operation being the usual matrix multiplication and where $\{\sigma^{\alpha_j}_j\}$ are the Pauli operators at site $j$. The group $\mathcal{P}_N$ has $4^{N+1}$ elements, half of these are Hermitian (and the other half anti-Hermitian). A \textit{Pauli stabilizer group} $\mathcal{S}$ is an Abelian subgroup of $\mathcal{P}_N$ such that $-I \not\in \mathcal{S}$. By excluding $-I$ from $\mathcal{S}$, we guarantee that all the elements of $s \in \mathcal{S}$ are Hermitian and  {can stabilize a subspace with eigenvalue $(+1)$}. The subspace $\mathcal{C}$ of vectors $|\psi \rangle$ satisfying $\forall s \in \mathcal{S}, s|\psi \rangle = |\psi \rangle$ is called \textit{the subspace stabilized by the group} $\mathcal{S}$. To define the stabilizer Hamiltonian associated to $\mathcal{S}$ we need to introduce the generating set $\cG \subset \cS$ of the stabilizer group. Recall that every Pauli stabilizer group may be written as the group generated by $M \leq N$  hermitian and commuting Pauli operators $\mathcal{G}=\left \{ g_1, \ldots , g_M \right \}$. If the generators are all independent the corresponding stabilized subspace $\mathcal{C}$ is $2^{N-M}$ dimensional.

A \textit{stabilizer Hamiltonian }$H$ is a Hermitian operator on $\bC^{2^N}$ which may be written with constants $J_k\geq0$ as 
\be\label{stabHamiltonian} 
H= - \sum_{k=1}^M J_k g_k
\ee 
of stabilizer operators $g_k \in \cG$. Note that the Hamiltonian is only defined in terms of the generating set $\cG$ and not the group $\cS$. The excitations therefore depend on the  choice of generators. The ground space, can be identified with the stabilized subspace $\mathcal{C}$. Stabilizer Hamiltonians are frustration free and can be diagonalized easily because all $g_k$ commute. Nevertheless they can exhibit exotic quantum behavior, such as topological order. For a review and a class of example Hamiltonians, the reader is referred to  {Ref.} \cite{Kitaev}.\\


\textit{The Davies generator}: We model the thermalizing dynamics of the stabilizer Hamiltonians by weak (or singular) local couplings to a large memoryless bath in thermal equilibrium. In this setting, there is a standard procedure for deriving a Markovian master equation describing the dissipative action of the bath on the system \cite{Davies}. Without loss of generality, we can assume that the system-bath interaction Hamiltonian is given by $H_{int} =  \sum_{k, \alpha_k}S_{\alpha_k}\otimes B_{\alpha_k}$, where $ B_{\alpha_k}$ are some set of operators acting on the bath and $S_{\alpha_k}\equiv\sigma^{\alpha_k}_k$ are the system coupling operators acting on a single site as Pauli operators, which span the full local matrix algebra. The master equation that is obtained from this procedure is   {$\cL(f)=i[H_{\rm eff},f]+\cD(f)$, where $f$ is an observable, and} $H_{\rm eff}$ is an effective Hamiltonian which does not contribute to the   {spectrum} of $\cL$.  The map $\cD$ describes the dissipative part of the evolution, and is given by 

\be\label{DaviesGenerator}
\cD(f) = \sum_{\alpha_k,\omega}h_{\alpha_k}(\omega) \left(S_{\alpha_k}^{\omega \dag} f S^{\omega}_{\alpha_k} -\half \{S^{\omega \dag}_{\alpha_k} S_{\alpha_k}^\omega,f\}\right),
\ee
where $\omega=E_i-E_j$ are the Bohr frequencies of $H$ with eigenvalues $E_i$. The rate $h_{\alpha_k}(\omega)$, with $h_{\alpha_k}(-\omega) = e^{-\beta \omega}h_{\alpha_k}(\omega)$ is determined by the bath auto-correlation function and encodes the dependence on $\beta$. We assume that  $h_{\min} = \min_{\alpha, \omega^\alpha} h^\alpha(\omega^\alpha)$ is the smallest transition rate over all realized Bohr frequencies. The jump operators  $S_{\alpha_k}^\omega$ are the Fourier coefficients of the time evolved system coupling operators $\exp(i H t) S_{\alpha_k} \exp(-i H t) = \sum_\omega e^{-i\omega t}S_{\alpha_k}^\omega$. For each $(\omega,\alpha_k)$ the  action of the jump operators  can be understood as mapping eigenvectors of $H$ with energy $E$ to eigenvectors of $H$ with energy $E+\omega$, and transfer energy $\omega$ from the system to the bath and back. The generator satisfies quantum detailed balance with respect to the Gibbs state $\rho$ of the Hamiltonian $H$ \cite{Davies}. This ensures that  $\rho$ is a fixed point and furthermore that the spectrum of $\cL$ is real.\\


\textit{The low temperature regime}: It has been shown  \cite{GibbsSamplers} that  Davies generators of bounded one dimensional lattice systems are always gapped. The proof hinges on an equivalence theorem between the relaxation time of Gibbs samplers and the correlation properties of the Gibbs state. Hence, in 1D   {there exists} a correspondence between the static and the dynamical critical behavior. We want to identify properties of the Hamiltonian that lead to a short mixing time independent of temperature,   { in dimensions larger than one. The following can be shown}: the central quantity which determines the spectral gap at low temperature is a slight generalization of the well known energy barrier for logical operators \cite{NoGo1}.

In order to define the \textit{generalized energy barrier} $\bar{\epsilon}$, we need to choose an enumeration $\Gamma$ of the lattice points, by which every Pauli operator is constructed by single qubit operations. For any $\eta \in {\cal P}_N$, we construct a path in Pauli space by setting $\eta_0 = \1$ and defining at every step $l \in \Gamma$ the Pauli operator $\eta_l$ to correspond to $\eta$ up to site $t$ and identity on the rest, so that the full Pauli is $\eta = \eta_{l_*}$ with $l_*$ the largest element in $\Gamma$. We define the {\it reduced generating set} of the Pauli operator $\eta$ by   ${\cal G}_\eta = \{ g \in {\cal G} \; ; \; [g,\eta] = 0 \}$, i.e. all terms in the Hamiltonian that commute with $\eta$. If we set $J = \max_k |J_k|$ the {\it energy penalty} of $\eta$ w.r.t. to the ordering $\Gamma$ is defined as
\be
	\epsilon_\Gamma(\eta) =2J \max_{l \in \Gamma}  \#\left \{ g \in {\cal G}_\eta \; ; \; [\eta_l , g] \neq 0 \right \}. 
\ee 
The constant that determines the spectral gap is now obtained by considering the highest energy penalty of any Pauli string with respect to the best labeling of lattice sites. That is we define
\be\label{genEnBar}
	\overline{\epsilon} = \min_{\Gamma} \max_{\eta \in \cP_N} \epsilon_\Gamma(\eta).
\ee
This quantity gives rise to the following spectral gap bound of the Davies generator. 
\be\label{lowTcBND}
	\lambda_D \geq \frac{h_{\min}}{4 l_*} e^{-2\beta \overline{\epsilon}}.
\ee
where $l_*= |\Gamma|$ denotes the length of the longest Pauli path

This result was first shown in \cite{Kristan1}. A bound to $\lambda_D$ is obtained by a direct evaluation of the Poincare inequality (Supp. Mat. Eqn. (\ref{PoncareInequality})) for the Davies generator. The   {gap} can be related to the energy barrier   {using methods} developed in   {Refs. \cite{JerrumSinclair,Kristan1}}.

\begin{figure}
\centering
  \includegraphics[scale=0.68]{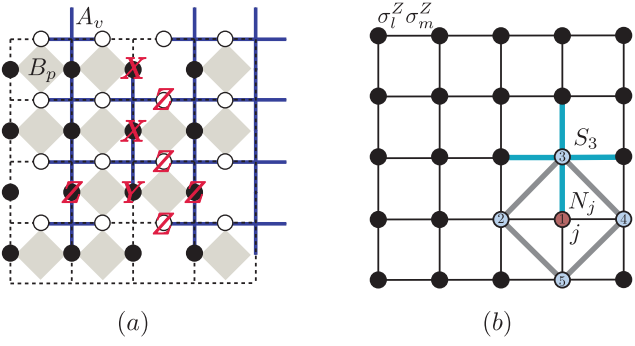}
    \caption{\label{fig1}(color online) For illustration purposes, we focus on the two simplest example systems, the toric code (TC) fig. (a), to explain the {\it low temperature bound} and the Ising model (IM) fig. (b) for the {\it high temperature bound}. Figure (a): The TC \cite{Kitaev} describes a lattice system of $N$ spins at every link on the lattice indicated by black / white dots. The Hamiltonian $H = -J\sum_v A_V -J\sum_pB_p$ is given as a sum of (blue) vertex terms $A_v = \sigma^Z_{v_1}\sigma^Z_{v_3}\sigma^Z_{v_3}\sigma^Z_{v_4}$ for every vertex $v$ and (grey) plaquette terms $B_p = \sigma^X_{p_1}\sigma^X_{p_2}\sigma^X_{p_3}\sigma^X_{p_4}$ for every plaquette $p$, so that $\cG= \{A_v,B_p\}_{v,p}$. We have depicted an example Pauli $\eta \in \cP_N$ by (red) $X,Y,Z$'s placed on the links. The corresponding reduced set $\cG_{\eta}$ has also been depicted by removing the appropriate (blue) vertex and (grey) plaquette  terms from $\cG$.     Figure (b): The IM is understood as a stabilizer   {Hamiltonian} where the qubits reside on the vertices  {:} $H = -J\sum_{\avr{ij}}\sigma^Z_{i}\sigma^Z_{j}$,so that  $\cG = \{\sigma^Z_i\sigma^Z_j\}_{i \sim j}$. We have depicted the set of qubits in $N_j$ for some $j$ as a grey rhombus and chose a clockwise order of the qubits. We have $|N_j| = 5$ and $S_j = \{\sigma^Z_{l}\sigma^Z_{j}\}_{l \in N_j\backslash\{j\}}$ so that $|S_j| = 4$ and furthermore $|S_l \cap S_j| = 1$. We can evaluate Eqn. (\ref{easybound}) as $\kappa(\beta) = 120 e^{8\beta J}(e^{2\beta J} -1)$ from which  $(\beta_*J)^{-1} \sim 249$. }
\end{figure}

The bound   {in Eqn. (\ref{lowTcBND})}  is stated in terms of the optimal choice of $\Gamma$. However, any non optimal ordering also gives rise to a valid bound. The quality of the bound strongly depends on the choice for $\Gamma$. For the 2D toric code, for example, a good choice gives rise to $\overline{\epsilon} \leq 2J$, whereas a poor choice can easily lead to an exponentially small bound on $\lambda_D$ with $\overline{\epsilon} \leq \cO(\sqrt{N})$. For a suitable ordering of the lattice sites $\Gamma$, one typically has that the longest path $l_* = \cO(N)$.   {We expect} that the dependence on $l_*$  in Eqn. (\ref{lowTcBND})  is an artifact of the derivation, and that the true bound should rather be $\lambda_D \geq \cO(h_{\min}e^{-2\beta \overline{\epsilon}})$. In particular in the limit $\beta \raw 0$ Eqn. (\ref{lowTcBND}) predicts $\lambda = \cO(N^{-1})$, whereas the true value of the gap is known to be $\lambda = \cO(1)$ \cite{GibbsSamplers}. If the Hamiltonian is not protected by an extensive generalized energy barrier, i.e. we have $\overline{\epsilon} = \mbox{const}$, the system thermalizes in a time that scales as $t_{mix} =\cO(N^2)$ since $\|H\| = \cO(N)$ for local stabilizer Hamiltonians.  

We now consider a simple example of $\epsilon_\Gamma(\eta)$ for the situation illustrated in Fig.~\ref{fig1}(a).  We choose an ordering $\Gamma$ where all lattice points are traversed twice as to first build up the $Z$ and then the $X$ factors of $\eta$. The $Y$ Pauli's are read as first applying $Z$ and then $X$ on the same site. We  have that $l_* = |\Gamma| = 2N$. The qubits are ordered so that for $Z$ factors we start to traverse the white qubits in each column (top to bottom) moving to the next column on  the right. Then we traverse each row of black dots (left to right) and
     move to the next lower row. For the $X$-factors the roles are reversed.  We first order the black qubits in each column (top to bottom) and then traverse the rows of white qubits 
     (left to right). The ordering fixes the path $\eta_t$ for any Pauli matrix $\eta \in \cP_N$. This amounts to decomposing every $\eta$ into small patches of incomplete "logical operators" 
     \cite{NoGo1}. We see that along this path at most one $g_k \in \cG_\eta$ is violated so that $\epsilon_\Gamma(\eta) \leq 2J$, for all $\eta$. This implies a lower bound $\lambda_D 
     \geq (8N)^{-1}h_{\min}e^{-4\beta J}$. Had we reversed the ordering $\Gamma$ for $X$ and $Z$ the bound would be exponentially small.   {The generalized energy barrier $\overline{\epsilon}$ is related to the standard energy barrier of logical operators \cite{NoGo1} when the lattice can be decomposed into the support of equivalent minimal logical operators. Then the argument above can be generalized and $\overline{\epsilon}$ corresponds to the largest energy barrier of the canonical logical operators.} \\

\textit{The high temperature regime}: It is expected that the low energy excitations do not determine the equilibration times at sufficiently high temperature, since this behavior is typically associated with the disordered phase of the model. In Ref. \cite{GibbsSamplers}, it was shown that there must exist a transition temperature above which the spectral gap of the Liouvillian is constant in the system size. From physical intuition one would expect that the transition temperature that gives rise to a constant gap in the dynamics should coincide with the static critical temperature of the Gibbs state of the spin Hamiltonian \cite{LocalityTemp}. That this is the case has only been shown in specific cases \cite{Martinelli,StrookZiggy,GibbsSamplers} and it is not known whether it holds for general quantum spin models. We therefore only discuss the dynamical critical temperature for which rapid mixing sets in here. Below we give an explicit lower bound on this temperature. 

The bounds only depend on the coordination properties of the Hamiltonian and can be evaluated by considering the local neighborhood of single spins. The neighborhood $N_j$ of spin $j$ is defined as the set of qubits $l \in N_j$ that share a stabilizer generator, so that $ l , j \in \mbox{supp}(g)$ for some $g \in \cG$. This set essentially coincides with the support of all the $S_{\alpha_j}^w$. We are free to choose an ordering of the qubits in $N_j$  so that   {$j = 1$} is the smallest element in this order. Furthermore we define the set $S_l \subset \cG$ as the set of generators that have support on site $l$, i.e. $g \in S_l$ if $l \in \mbox{supp}(g)$. In Proposition (Sup. Mat. Prop. \ref{kappaBnd}) we prove a lower bound on a constant $\kappa$ that can be simplified to 

\be \label{easybound}
\kappa(\beta) = \max_j 3(|N_j| - 1) e^{2 \beta J |S_j|} \sum_{l \in N_j} \sum_{k > l}(e^{2 \beta J| S_j \cap S_k |} -1).
\ee

Here  $J=\max_{k \in N_j} J_k$ is the largest coupling constant. The bounds on temperature and gap are obtained from the following theorem.

\begin{theorem}\label{Thm:highTc}
Above a threshold temperature $\beta_*$, the gap of $\cL$ is independent of the system size. We get for   {any $\beta<\beta_*:=\max\{ \beta |\kappa(\beta) <  1\}$}, that 
$ \lambda_D \geq \half h_{\min}e^{-2\beta S_*} \left(1 - \kappa(\beta)\right)$, with $S_* = \min_j |S_j|$.
\end{theorem} 

\proof{ The proof follows closely ideas developed in   {Ref.} \cite{ZiggyFirst}. We defer the technical calculations of constants to the appendix, and only present the main steps here. To start, we define the so-called \textit{heat-bath generator}: $\cQ^*(\sigma)=\sum_j\cQ^*_j(\sigma)\equiv\sum_j \bE_j^*(\sigma) -\sigma$, where $\bE^*(\sigma)=\gamma_j^\dag\ptr{j}{\sigma}\gamma_j$, and $\gamma=(\ptr{j}{e^{-\beta H}})^{1/2}e^{-\beta H/2}$, here $\ptr{k}{f}$ denotes the normalized partial trace and corresponds to a single qubit twirl.   { $\cQ$ is a proper Liouvillian}, and by a comparison theorem (Supp. Mat. Lemma \ref{local_comp}) we can show that the gap of the Davies generator can be lower bounded by the gap $\lambda_Q$ of $\cQ$ as: 
 \be\label{DaviesLowComp}
  \lambda_D\geq R \lambda_Q, 
 \ee 
where we have the bound $R \geq \half h_{\min} e^{-2\beta S_*}$, which is system size independent. Hence, it suffices to lower bound the gap of the heat-bath generator, to get a lower bound on the gap of the Davies generator. The spectral gap of $\cQ$ can be bounded by a generalization of the Aizenman and Holley \cite{aizenman1987rapid} conditions to quantum spin systems \cite{ZiggyFirst}. 

Let $T_t=e^{t\cQ}$ be the heat-bath semigroup in the Heisenberg picture (i.e. acting on observables), and define the discrete gradient: $\partial_k(f)=f-\ptr{k}{f}$. Note that $\partial_k \cQ_k(f)=\partial_k(f)$, since $\partial_k \bE_k(f)=0$. Therefore, 
\be 
\frac{d}{ds}\partial_k T_s(f)=-\partial_k T_s(f)+\sum_{j\neq k}\cQ_j \partial_k (f)+\sum_{j\neq k}[\partial_k,\cQ_j]T_s(f)\nonumber
\ee
If we define $T^{(k)}_t=\exp(t\sum_{j\neq k}\cQ_j)$, we have 
\be \label{eqn:bdn3}
\frac{d}{ds}(e^sT^{(k)}_{t-s}\partial_kT_s(f))=\sum_{j\neq k}e^s T_{t-s}[\partial_k,\cQ_j]T_s(f)
\ee
Now, integrating from $0$ to $t$ on both sides of Eqn. (\ref{eqn:bdn3}) and using the fact that the semigroup $T_t$ is contractive (i.e. $||T_t(f)||\leq||f||$, for all $f$), we get
\be 
||\partial_k T_t(f)||\leq e^{-t}||\partial_k f||+\sum_{j\neq k}\int_0^t e^{-(t-s)}||[\partial_k,\cQ_j]T_s(f)||\nonumber
\ee
The trick now is to bound the last term point wise using Lemma (Supp. Mat. Lemma \ref{bound-on-a}), which claims that there exist a set of positive constants $a_{jk}^l\geq0$ such that  
\be 
||[\partial_k,\cQ_j](f)||\leq \sum_l a_{j,k}^l ||\partial_l(f)||,
\ee
with $\sum_k \sum_{j\neq k} a_{j,k}^l \leq \kappa$, where $\kappa<\infty$ for any $l,k$. Then we get:
\be
|||T_t(f)||| \leq e^{-t}|||f|||+\kappa\int_0^t e^{-(t-s)}|||T_s(f)|||,
\ee 
where $|||f||||=\sum_k ||\partial_k f||$ is the so-called \textit{oscillator norm} \cite{ZiggyFirst,ZiggyOther,Werner}.  
Iterating this equation yields $|||T_t(f)|||\leq e^{-(1-\kappa)t}|||f|||$. A telescopic sum argument \cite{ZiggyFirst} can then be applied to obtain 
\be
 ||T_t(f)-\tr{\rho f}||\leq e^{-(1-\kappa)t}|||f|||,\label{ergodicity}
\ee 
where $\rho$ is the Gibbs state of $H$. Clearly, whenever $\kappa <1$ we get a non-trivial bound. Invoking properties of $\bL_p$ norms (Supp. Mat. Lemma \ref{Gap-erg}), Eqn. (\ref{ergodicity}) implies that the spectral gap of $\cQ$ is lower bounded by $\lambda_Q \geq 1-\kappa$. The final step in the proof consists in finding an upper bound on $\kappa$. We obtain the bound Eqn. (\ref{easybound}) from (Sup. Mat. Prop. \ref{kappaBnd}) after additional simplifications, so that applying Eqn. (\ref{DaviesLowComp}) yields the result. \qed}

\vspace{0.2cm}
Typically we will find that $\beta_* \ll 1$ so that we can consider the expansion of $\kappa$ and only focus on the first order term to obtain a rough estimate of $\beta_*$. We have that $\kappa_j \approx 6\beta J (|N_j|-1)\sum_{l \in N_j}\sum_{k \geq l } | S_j \cap S_k |$, so that we can immediately solve $\kappa =1$ for $\beta_*$.

The bound we obtain yields the same functional dependence on the temperature as in the static case \cite{LocalityTemp}. However, it is \ {clearly not} optimal in practice. For instance, for the Ising model in two dimensions, our bound yields a dynamical critical temperature of $(\beta_*J)^{-1} \sim 249$, whereas the correct static critical temperature is $\sim2.27$.  For comparison, in Ref. \cite{LocalityTemp} a bound on the static critical temperature corresponded to a value of $\sim24.58$. Since the bound only depends on the local properties of the stabilizer Hamiltonian it can be evaluated easily, however this implies  that the long range aspects, needed for a tight bound on the critical temperature, are neglected. The fact that for $\beta \leq \beta_*$ we have that $\lambda = \mbox{const}$ implies again with $\|H\| = \cO(N)$ for local stabilizer Hamiltonians that $t_{mix} = \cO(N)$.\\

\textit{Conclusions}: 
We have presented a complete characterization of the thermalization time for stabilizer Hamiltonians. By now it is known \cite{Michnicki,Haah} that the existence of an energy barrier is not sufficient for the thermal stability of a SCQM. However, the low temperature result shows that at least for Pauli stabilizer Hamiltonians {\it the existence of an energy barrier $\overline{\epsilon}$ is a necessary condition}. This implies in particular, that mechanism such as the ones analyzed in \cite{Stark,Pachos}, where random perturbations in the couplings suppress the coherent propagation of excitations, cannot protect against thermal errors. Furthermore, we find an explicit bound on the critical temperature above which the gap of the generator is constant in the system size. The bound confirms the intuition that at sufficiently high temperature the particular low energy properties of the model do not influence the thermalization dynamics and only local properties of the model matter.\\

{\it Acknowledgements:} We  thank F.\ Pastawski and F.\ Brandao for helpful discussions.   {This work was supported by the Carlsbergfond, the Villum foundation, the Humboldt foundation,} the Institute for Quantum Information and Matter, a NSF Physics Frontiers Center with support of the Gordon and Betty Moore Foundation (Grants No. PHY-0803371 and PHY-1125565)


%

\begin{widetext}
\section{Supplemental Material}
\begin{appendix}

\subsection{Stabilizer Heat-bath and Davies generators}

In this subsection, we will analyze the Heat-bath Liouvillian $\cQ$ defined in the proof of Theorem \ref{Thm:highTc} and derive a more explicit representation for stabilizer Hamiltonians.
Moreover, we will introduce notation that will facilitate the derivation of the constants discussed in the main body of the text. Since the terms in the Hamiltonian 
\be 
H = -\sum_k J_k g_k
\ee 
commute, it can be diagonalized in the common eigenbasis of all $\{g_k\}$.  The individual $g_k$  have eigenvalue $\pm 1$, so the projectors onto their eigen-subspaces are $ \Pi_k(b_k):=\half (\1 + (-1)^{b_k} g_k)$. Here the choice $b_k = 0,1$ denotes the projector onto the local subspace with eigenvalue $(-1)^{b_k}$. Since all  $\{g_k\}$ commute, we can construct global projectors  by taking their products. Hence, eigenvalues and corresponding subspaces can be described by  an $M = |\cG|$ component bit string $b$.  That is, we have that for every $b \in \{0,1\}^M$, the eigenvalues are
\bq
\epsilon(b) := -\sum_k J_k (-1)^{b_k}, \Sp \mbox{with projectors} \Sp P(b):= \Pi_1(b_1)\Pi_2(b_2) ... \Pi_M(b_M).
\eq
With this at hand we can diagonalize the Hamiltonian immediately and obtain 
\be\label{spec-decomp}
H = \sum_b \epsilon(b) P(b). 
\ee 
The reader familiar with the stabilizer formalism will immediately recognize the bit string $b$ as the syndrome of the stabilizer code \cite{Gottesman} generated by $\cG$.  
The Gibbs state of $H$ can immediately be written as $ Z^{-1} e^{-\beta H} =   Z^{-1} \sum_b e^{-\beta \epsilon(b)}P(b)$.  For convenience we will also frequently write $\rho_a = Z^{-1}\exp(-\beta\epsilon(a))$, so that $\rho = Z^{-1} e^{-\beta H} = \sum_a \rho_a P(a)$. \\ 

To analyze the action of the Davies or Heat-Bath generators,  it is necessary to understand how the Pauli operators $\cP_N$ act on the eigen projectors of the Hamiltonian. We therefore need to introduce some notation. Recall that the Pauli group is isomorphic to $\cP_N \simeq \bZ_2^{2N+2} = \{0,1\}^{2N+2}$, where two bits are needed to encode the phase  information. Since we work with the Pauli matrices as a basis we do not need to carry the phase information and can identify every Pauli operator up to a phase with a $2N$ - bit string $\alpha \in  \bZ_2^{2N}$. We write for this Pauli matrix  $\sigma(\alpha) \in \cP_N$. In particular we denote the single qubit Pauli operators acting on site $j$ by $\sigma(\alpha_j) = \sigma^\alpha_j$, where $\alpha_j = (0,0)_j,(0,1)_j,(1,0)_j,(1,1)_j$ corresponds to $\1,\sigma^x_j,\sigma^z_j,\sigma^y_j$ respectively. The full algebra is indexed by concatenating these bit strings from every site. Given the set $\cG$ of Pauli's which comprise the Hamiltonian $H$, we can now define a syndrome, or excitation to every Pauli $\sigma(\alpha)$ to which we will refer as $e(\alpha) \in \bZ^M$ which is a $M$-bit string defined as $e(\alpha) = (e_1(\alpha,)\ldots e_M(\alpha))$ defined by commutation properties $\sigma(\alpha) g_k  = (-1)^{e_k(\alpha)} g_k\sigma(\alpha)$ for every $k$. Hence the action of a Pauli operator on a projector $P(a)$ can be understood as
\be \label{commu-proj}
	\sigma(\alpha)P(a) = P(a \oplus e(\alpha)) \sigma(\alpha),
\ee
by direct observation. For notational purposes we will write the bit string $a^\alpha := a \oplus e(\alpha) $ so that $P(a \oplus e(\alpha)) \equiv P(a^\alpha)$ and $\epsilon(a \oplus e(\alpha))=\epsilon(a^\alpha)$ respectively. Furthermore, it is helpful to define the difference between two eigenvalues $\epsilon(a)$ and $\epsilon(a^\alpha)$, when one is obtained from the other by Pauli excitation $\sigma(\alpha)$. That is we will frequently consider the Bohr frequency
\be
	\omega^\alpha(a) = \epsilon(a) - \epsilon(a^\alpha) = -\sum_{k}2J_k(-1)^{a_k} e_k(\alpha),
\ee  
where by abuse of notation we read $e_k(\alpha) \in \bZ_2$ as a $\{0,1\}$-valued variable with standard addition in $\bR$. The purposes of the newly introduced notation is to encode the excitations of the stabilizers Hamiltonians in a compact manner that is model independent. This is approach was initiated in \cite{Kristan1} to obtain the low temperature bound in Eqn. (\ref{lowTcBND}) in the main body of the text. With this notation at hand, we are now in the position to analyze the action of the heat-bath and  Davies maps of $H$ on the matrix algebra $\bC[\cP_N] \simeq \cM_{2^N}(\bC)$.\\

\paragraph{\bf The heat-bath generator:}We now analyze the heat-bath generator $\cQ = \sum_{j} \cQ_j$ that was introduced for the proof of Theorem \ref{Thm:highTc}. The Heat-bath Liouvillian can be understood as a quantum extension of the classical Glauber dynamics.  Recall that $\cQ_j(f) = \bE_j(f) - f$, with $\bE_j(f) = \ptr{j}{\gamma_j^\dag f \gamma_j}$, and $\gamma_j = \ptr{j}{e^{-\beta H}}^{-1/2} e^{-\beta H/2}$. The conditional expectation $\bE$ is a quantum channel (cot map). We will write the generator  in Lindblad form as this will simplify its spectral analysis. For a qubit lattice system $\Lambda$, the {\it heat-bath Liouvillian} $\cQ$ can be written in Lindblad form as 
\be\label{eqn:heatbath}
\cQ(f) \assign  \sum_{j \in \Lambda}\sum_{\alpha_j}  A_j^{\alpha \dag } f A_j^{\alpha} - \frac{1}{2} \{ f, A_j^{\alpha \dag} A_j^{\alpha}\}
\ee
with jump operators $A_j^{\alpha}  = \sum_{b} \half G_j(b) P(b) \sigma(\alpha_j)$ where $\sigma(\alpha_j)$ are the single qubit Pauli matrices acting on site $j$ and 
\be\label{def:G-func}
G_j(b) \equiv \left( \frac{1}{4} \sum_{\alpha_j} e^{\beta\omega^{\alpha_j}(a)} \right)^{-1/2}.
\ee

\vspace{0.5cm}
Observe that the normalized partial trace $\ptr{j}{\cdot}$ over a single site $j$ (local depolarizing map) can be expressed as a sum over Pauli matrices $\sigma(\alpha_j)=\{\1,\sigma^X_j,\sigma^Y_j,\sigma^Z_j\}$ as 
\begin{equation} 
\ptr{j}{f}=\frac{1}{4} \sum_{\alpha_j} \sigma(\alpha_j) f \sigma(\alpha_j).
\end{equation}

We will now turn to expressing $\cQ$ in a specific basis of operators. First observe that,
 
\begin{align}
\ptr{j}{e^{-\beta H}}=& \frac{1}{4}\sum_{\alpha_j} \sum_b e^{-\beta \epsilon(b)} \sigma(\alpha_j) P(b)\sigma(\alpha_j) = \frac{1}{4}\sum_{\alpha_j} \sum_b e^{-\beta \epsilon(b^{\alpha_j})} P(b).
\end{align}

Where the last equality follows from Eqn. (\ref{commu-proj}) by adding $e(\alpha_j)$ to all $b$  and relabeling. We then get

\begin{align}
\gamma_j =& (\ptr{j}{e^{-\beta H}})^{-1/2} e^{-\half \beta H} \no 
=&\left(\frac{1}{4}  \sum_{\alpha_j} \sum_b e^{-\beta \epsilon(b^{\alpha_j})} P(b)\right)^{-1/2}\left( \sum_{b'} e^{-\half\beta \epsilon(b')} P(b')\right)\no
=& \left(\sum_b  \left( \sum_{\alpha_j}  \frac{1}{4} e^{-\beta \epsilon(b^{\alpha_j})} \right)^{-1/2} P(b) \right) \left( \sum_{b'} e^{-\half\beta \epsilon(b')} P(b')\right)\no
=&  \sum_b \left( \frac{1}{4} \sum_{\alpha_j} e^{-\beta \epsilon(b^{\alpha_j})} \right)^{-1/2}e^{-\half \beta \epsilon(b)}P(b) \equiv  \sum_b G_j(b)P(b).
\end{align}

We have defined $G_j(b)$ in (\ref{def:G-func}) and made use of the definition $\omega^{\alpha_j}(b) = \epsilon(b) - \epsilon(b^{\alpha_j})$. Recall that the Liovilian is given by, $\cQ_j(f) = \ptr{j}{\gamma_j^\dag f \gamma_j} - f$ so we can write the trace as the twirl over the single site Pauli group as explained above. Hence, we may choose the Lindblad operators 
\be \label{eqn:LB-heat}
A_j^{\alpha\dag} = \frac{1}{{2}}\sigma_j^\alpha \gamma_j^\dag = \sum_{b} \half G_j(b) \sigma^\alpha_j P(b).
\ee
Moreover, note that since for any operator $f$ we have that $\ptr{j}{f}$ commutes with any other operator that is only supported on site $j$ we obtain
\bq
&&\sum_{\alpha} A_j^{\alpha\dag}A_j^{\alpha} = \frac{1}{4} \sum_{\alpha} \sigma_j^\alpha \gamma_j^\dag \gamma_j \sigma_j^\alpha \no 
&& = \ptr{j}{e^{-\beta H}}^{-1/2} \ptr{j}{e^{-\beta H}} \ptr{j}{e^{-\beta H}}^{-1/2} = \1.
\eq
This leads in the end to the decomposition of the generator $\cQ_j$ into Lindblad operators as given above.

\vspace{0.5cm}
\paragraph{\bf Davies Generator:} For completeness we also briefly discuss the particular form of the Davies generators for the spectral decomposition of $H$ introduced above. Recall that the Liouvillian is obtained by weakly coupling the system  $H$  to a thermal bath  $\sigma_B = Z^{-1}\exp(-\beta H_B)$, where $H_B$ is the bath Hamiltonian. The system-bath interaction is given by  $H_{int} = \sum_{\alpha} S_{\alpha} \otimes B_{\alpha}$  where $S_\alpha$ are local Hermitian operators only acting on the system, that couple to the bath operators $B_\alpha$. It can be shown \cite{Davies} that in the weak coupling limit the reduced evolution on the system is described by a Master equation of the form $\partial_t \rho(t) = -i[H_{\rm eff},\rho] + \cD^*(\rho)$. Here $H_{\rm eff}$ is the effective, lamb shifted, system Hamiltonian. This Hamiltonian does not contribute to the spectral gap of the generator \cite{Kristan2} and we will therefore be ignore it in the following.\\

We will now consider the {\it Davies generators} and express them for  commuting Pauli Hamiltonians. Recall the definition of Davies generator as given in Eqn. (\ref{DaviesGenerator})
\bq
	\cD(f) = \sum_{\omega, \alpha_j} \cD_{{\alpha_j} ; \omega} (f),
\eq
where the individual terms are
\be\label{eqn:daviesmap}
  \cD_{{\alpha_j} ; \omega} (f) = h_{\alpha_j}(\omega) \left( S^{\omega \dag}_{\alpha_j} f S^{\omega }_{\alpha_j}  - \half\left\{S^{\omega \dag}_{\alpha_j} S^{\omega }_{\alpha_j} ,f\right\} \right)
\ee
We know that due to the KMS condition, the Fourier transform of the  bath auto-correlation function satisfies $h^{\alpha_j}(\omega)=e^{-\omega \beta}h^{\alpha_j}(-\omega)$. To state a bound that is independent of the specific Bath properties, we assume that the transition rates are bounded from below and above by a possibly temperature dependent constant independent of the system size: $h_{\max} \geq h^\alpha_j(\omega) \geq h_{\min} > 0$. Moreover, recall that 
 \be
	e^{iH t} S^\alpha_j e^{ - iH t} = \sum_{\omega} S^{\omega }_{\alpha_j} e^{i \omega t}.
\ee 
We choose the local perturbations on site $j$ simply as all the single qubit Pauli operators $S^\alpha_j = \sigma(\alpha_j)$. Since $H$ is a stabilizer Hamiltonian, we can write its time evolution as $\exp(i Ht) = \sum_b \exp(i \epsilon(b) t) P(b)$ by means of the spectral decomposition in Eqn. (\ref{spec-decomp}) From which the individual $S_j^{\alpha} ( \omega)$ can be read off directly
\be\label{Jump_opDavies}
S^{\omega }_{\alpha_j} \assign  \sum_b \delta [ \omega - \omega^{\alpha_j}(b)]   \sigma(\alpha_j) P ( b),
\ee
where he have defined the following $\delta$ function
\bq
\delta[x] = \left \{ \begin{array}{l} 1 \Sp : \Sp \mbox{for} \Sp x = 0 \no 0 \Sp : \Sp \mbox{else}. \end{array}\right.
\eq

\section{Comparing spectral gaps}

It is sometimes possible to bound the convergence behavior of one generator by that of another.  In our case this has the following consequence. Suppose we obtain a bound on the spectral gap $\lambda_Q$ of the Heat bath dynamics, as was done in the proof of Theorem \ref{Thm:highTc}, we can then use this bound to obtain a bound on the spectral gap of the Davies generator by means of a comparison technique. The general argument for such a comparison of spectral gaps can be made along the following lines. The Poincare inequality \cite{chi2} gives rise to a variational characterization of the spectral gap of a Lindbladian $\cL$.
\be\label{PoncareInequality}
	\lambda \Var_{\rho}(f) \leq \cE(f) \Sp\Sp \forall f\in \bC^{2^N},
\ee
where the two forms are defined as 
\be
\Var_{\rho}(f) = \tr{\rho^\half f^\dag \rho^\half f} - |\tr{\rho f}|^2 \Sp \mbox{and} \Sp \cE(f) = -\tr{\rho^\half f^\dag \rho^\half \cL(f)}.
\ee
We refer to $\cE(f)$ as the Dirichlet from and to $\Var_\rho(f)$ as the variance.  A direct evaluation of this inequality was carried out in \cite{Kristan1} and yields the inequality on the gap as stated in Eqn. (\ref{lowTcBND}). For the high temperature bound we use this inequality as an intermediate step. If we can find a constant $\tau$ such that $\cE^Q(f) \leq \tau \cE^D(f)$, where $\cE^Q(f)$ and $\cE^D(f)$ are the Dirichlet forms for $\cQ$ and $\cD$ respectively, we can infer a lower bound on the spectral gap $\lambda_D$ of the Davies generator in terms of $\lambda_Q$ from a simple chain of inequalities. 
\begin{lemma}
Let $\lambda_Q$ denote the spectral gap of the  heat bath Liovillian $\cQ$ (Eqn. (\ref{eqn:heatbath})) associated to the commuting Pauli Hamiltonian $H = - \sum_kJ_k g_k$, then the spectral gap $\lambda_D$ of the corresponding Davies generator $\cD$ (Eqn. (\ref{eqn:daviesmap})) can be bounded by
\be
	\lambda_D \geq R \lambda_Q.
\ee
The constant $R$ is given by
\be
	R  \geq  \;  \half h_{\min}e^{-2\beta S_*},
\ee
where in the lower bound we have defined $h_{\min} = \min_{\omega^{\alpha_j}(a)}h^{\alpha}(\omega^{\alpha_j}(a))$ and $S_* = \max_{j \in \Lambda} |S_j|$.
\end{lemma}

\proof{ The Poincare inequality, Eqn. (\ref{PoncareInequality}), gives a variational characterization of the spectral gap. From the inequality we see that the gap is the largest constant $\lambda$ that satisfies $\lambda \Var_\rho(f) \leq \cE(f)$ for all $f \in \bC^{2^N}$. Note that both the heat bath generator $\cQ$ as well as the Davies generator $\cD$ have the same fixed point $\rho = Z^{-1}\exp(-\beta H)$. This means that we can use the same variance for both inequalities. Hence by virtue of $\lambda_Q \Var_\rho(f) \leq \cE^Q(f)$ and applying lemma \ref{local_comp}, we get
\be
	\lambda_Q \Var_\rho(f) \leq \cE^Q(f) \leq \tau \cE^D(f), \Sp \forall f\in\bC^{2^N}
\ee
so that have $\lambda_D \geq R \lambda_Q$ with $R = \tau^{-1}$. We can directly bound $\tau$ using: $\max_{a,j} G_j(a) = 4^{-1} \max_{a,j} \sum_{\alpha_j} e^{\beta \omega^{\alpha_j}(a)} \leq e^{2\beta S_* }$ to obtain the stated bound on $R$. \qed}

\vspace{0.3cm}
Since both maps are very similar in their support, we will be able to compare the Dirichlet forms of Davies Generator and the Heat bath dynamics locally. That is we will proceed to split the 
forms up in ever smaller subsystems and identify the worst bound among these simpler systems. This will then lead to a direct comparison between the two forms.

\begin{lemma}\label{local_comp}
Let $\cE^Q(f) = -\tr{\rho^\half f^\dag \rho^\half \cQ(f)}$ denote the  Dirichlet form of the heat-bath generator and let $\cE^D(f) = -\tr{\rho^\half f^\dag \rho^\half \cD(f)}$ denote the Dirichlet form of the  Davies generator, for any $f\in\bC^{2^N}$.Then
\bq
	\cE^Q(f) \leq \tau \cE^D(f),
\eq
where 
\be 
\tau = \max_{j;a}  \; 2 \;\frac{G^2_j(a)}{h_{\min}}.
\ee
\end{lemma}

\proof{ The fact that we are considering commuting Hamiltonians ensures, that both the Davies generator as well as the thermal course graining map can be written as a sum of local generators that are supported only in the neighborhood of a single site. We shall write in the following 
\be
\cE^Q(f) = \sum_{j \in \Lambda} \cE^Q_j(f) \Sp \mbox{as well as} \Sp  \cE^D(f) = \sum_{j \in \Lambda} \cE^D_j(f),
\ee
with the local forms $\cE^Q_j(f) = -\tr{\rho^\half f^\dag \rho^\half \cQ_j(f)}$ and $\cE^D_j(f) = -\tr{\rho^\half f^\dag \rho^\half \cD_j(f)}$, and $\Lambda$ is the full lattice. If we can find constants $\tau_j$ so that the local terms are bounded as $\cE^Q_j(f) \leq \tau_j \cE^D_j(f)$ we know that the choice $\tau = \max_j \tau_j$ will provide the desired bound. To find bounds on the local $\tau_j$ we follow an approach taken in Ref. \cite{Kristan2} and recast the problem into a semi definite inequality. We vectorize the $f \in \bC^{2^N}$, by the natural isomorphism that maps $f  \raw \oket{f} \in \bC^{4^N}$ and write both forms in terms of hermitian Matrices $\cE^Q_j(f) = \obra{f} \hat{\cE}_j^Q \oket{f}$ and $\cE^D_j(f) = \obra{f} \hat{\cE}_j^D \oket{f}$. The local number $\tau_j$ can now be bounded by the smallest number such that 
\be
	\tau_j \hat{\cE}_j^D - \hat{\cE}_j^Q \geq 0,
\ee
is positive semi-definite. We will see that $\hat{\cE}_j^D$ and $ \hat{\cE}_j^Q$ are very similar in form, so we will be able to find good bounds on $\tau_j$ easily.\\

We only derive the matrix $\hat{\cE}_j^Q$ because a similar derivation has already been performed in Ref. \cite{Kristan1} for the Davies generator. We will express the matrix entries of $\hat{\cE}_j^Q$ in terms the Pauli matrices normalized with respect to the Hilbert-Schmidt inner product.  We have that the action of $\hat{\cE}^Q_j(\sigma(\gamma)) = -\sqrt{\rho} \cQ_j(\sigma(\gamma)) \sqrt{\rho}$ defines the matrix $\hat{\cE}^Q_j$. We evaluate
\bq
 \hat{\cE}^Q_j(\sigma(\gamma)) =  \sum_{\alpha_j} \sqrt{\rho} \left( \frac{1}{2} \{ A_j^{\alpha \dag} A_j^{\alpha},\sigma(\gamma)\} -A_j^{\alpha \dag } \sigma(\gamma) A_j^{\alpha}\right) \sqrt{\rho},
\eq
by using the representation of $A^\alpha_j$ in Eqn. (\ref{eqn:LB-heat}), and by virtue of $\rho = \sum_a \rho_a P(a)$ we obtain

\bq
\hat{\cE}^Q_j(\sigma(\gamma)) = \frac{1}{4} \sum_{\alpha_j}\sum_{a,b,l,m} \; G_j(a)G_j(b)\sqrt{\rho_l\rho_m}\Big{(} \; \half \; \delta_{a,b} P(l)\left\{\sigma(\alpha_j)P(a)\sigma(\alpha_j),\sigma(\gamma)\right\}P(m)  \no  - P(l)\sigma(\alpha_j)P(a)\sigma(\gamma)P(b)\sigma(\alpha_j)P(m)\Big{)}
\eq

If we now make use of the commutation relations Eqn. (\ref{commu-proj}) between $\sigma(\eta)$ and $P(a)$ we have that
\bq\label{DirichletHB}
\hat{\cE}^Q_j(\sigma(\gamma)) = \frac{1}{4}\sum_{\alpha_j} \sum_{a} \left( \half\left(G^2_j(a^{\alpha_j}) + G^2_j(a^{\alpha_j,\gamma})\right) 
- G_j(a^{\alpha_j})G_j(a^{\alpha_j,\gamma}) \; \theta_{{\alpha_j},\gamma}\right)\sqrt{\rho_a \rho_{a^\gamma}}  \;P(a) \sigma(\gamma). 
\eq
Here we have defined the phase $\theta_{\alpha_j,\gamma} = \pm1$ depending on wether the Pauli matrices commute or anti-commute $\sigma(\alpha_j)\sigma(\gamma) = \theta_{\alpha_j,\gamma} \sigma(\gamma)\sigma(\alpha_j)$.\\

In this form it is possible to read off the matrix $\hat{\cE}^Q_j$ in the Pauli basis. As discussed previously, there is a correspondence between $\bZ^{2N+2}$ and ${\cal P}_N$.
We only care about the Pauli matrices and not the phase information of the group. We associate to every Pauli matrix, labeled by $\gamma \in \bZ_2^{2N}$, a vector 
in $\bC^{4^N}$ using the state $\oket{\Omega}  = \sum_{k \in \bZ_2^{N}} \frac{1}{\sqrt{2^N}} \ket{k,k}$, through
\be
	\oket{\gamma}  =  \sigma(\gamma)\otimes \1 \oket{\Omega}.
\ee 
Moreover, note that we can express the projectors as $P(a) = 2^{-M} \sum_x (-1)^{\avr{x,a}} g_1^{x_1}\ldots g_M^{x_M}$. We can define a dual basis for each coset of the Stabilizer group $\cS = \avr{\cG}$ in the full Pauli group $\cP_{N}$. Following \cite{Kristan2} we denote the coset Basis elements by $\oket{a_{\gamma_0}}$ and define them by
\be\label{coset_dual}
\oket{a_{\gamma_0}} = \frac{1}{2^{M/2}} \sum_{x \in \bZ_2^{M}} (-1)^{\avr{a,x}}  \left( g_1^{x_1}\ldots g_M^{x_M}\sigma(\gamma_0) \otimes \1\right) \oket{\Omega}.
\ee
Here $a$  corresponds to the syndrome of the generator set $a$, and $\gamma_0$ labels the representative of the coset. Following \cite{Kristan1}, we can express the matrix in this basis
and read off the coefficients from Eqn. (\ref{DirichletHB}) to obtain

\bq\label{DirichletMatrix}
\hat{\cE_j}^{Q} =  \sum_{ [\gamma_0] \in \cP_N \backslash \cS} \; \sum_{a , \alpha_j} \; A^{Q}_{\alpha_j,{\gamma_0}}(a) \oproj{a_{\gamma_0}} -  B^{Q}_{\alpha_j, \gamma_0}(a) \theta_{\alpha_j,{\gamma_0}} \oket{a_{\gamma_0}}\obra{a^\alpha_{\gamma_0}}.
\eq 

Here we have defined the following functions

\bq
A^{Q}_{\alpha,\gamma}(a) = \frac{1}{8}\left(G^2_j(a^\alpha) + G^2_j(a^{\alpha,\gamma})\right)\sqrt{\rho_a \rho_{a^\gamma}} \Sp \mbox{and} \Sp B^{Q}_{\alpha,\gamma}(a) = \frac{1}{4}G_j(a^\alpha)G_j(a^{\alpha,\gamma}) \sqrt{\rho_a \rho_{a^\gamma}}.
\eq

Note, that these functions are constant over the coset $[\gamma_0]$, i.e.  one can easily verify that if we choose a different representative $\gamma_1$ of the same coset we have that $A^{Q}_{\alpha,\gamma_1}(a) = A^{Q}_{\alpha,\gamma_0}(a)$ and $B^{Q}_{\alpha,\gamma_0}(a) = B^{Q}_{\alpha,\gamma_1}(a)$. Moreover the same holds for $  \oproj{a_{\gamma_0}} =  \oproj{a_{\gamma_1}}$ as well as for $\theta_{\alpha_j,{\gamma_0}} \oket{a_{\gamma_0}}\obra{a^\alpha_{\gamma_0}} = \theta_{\alpha_j,{\gamma_1}} \oket{a_{\gamma_1}}\obra{a^\alpha_{\gamma_1}}$. This can be verified directly from the definition in Eqn. (\ref{coset_dual}). Hence, the choice of representative is arbitrary.\\

Essentially  the same analysis for the {\it Davies generator} has already  been done in Ref. \cite{Kristan1} to obtain the corresponding Dirichlet matrix. It turns out that the 
matrix is of the same form and we only need to replace $A$ and $B$ in Eqn. (\ref{DirichletMatrix}) by the corresponding versions for the Davies generator, which are
\be
A^{D}_{\alpha,\gamma}(a) = \half\left(h_{aa}^{\alpha} + h_{a^{\gamma} a^{\gamma}}^{\alpha}  \right)\sqrt{\rho_a \rho_{a^{\gamma}}} 
\Sp \mbox{and} \Sp B^{D}_{\alpha,{\gamma}}(a) = h^{\alpha}_{a,a^{\gamma}} \sqrt{\rho_a \rho_{a^{\gamma}}},
\ee
where we have defined the following function $h_{ab}^\alpha = h^{\alpha}(\omega^\alpha(a))\delta[\omega^\alpha(a) - \omega^\alpha(b)]$, which arises from the definition of $\cD$ in Eqn.
(\ref{Jump_opDavies}) by taking the sum over the Bohr frequencies $\omega$. The Dirichlet matrix  therefore has the same decomposition over cosests $[\gamma_0]$, syndromes $a$ and single qubit Pauli's $\alpha_j$. This provides an easy approach to finding bounds on the support numbers $\tau_j$. \\

The dual vectors $\oket{a_{\gamma_0}}$ are orthogonal if they come from different cosets and thus the full Dirichlet matrix is a direct sum over the cosets $[\gamma_0]$. We can therefore write for the matrix 
\be
	\hat{\cE}_j^{Q/D} = \bigoplus_{[\gamma_0]} \hat{\cE}_{j,\gamma_0}^{Q/D},
\ee
where each of the matrices $\hat{\cE}_{j,\gamma_0}^{Q/D}$ can be decomposed into a sum of positive semi-definite two dimensional matrices. We can then support the two dimensional matrices from the heat-bath generator by those of the Davies generator, reducing the complexity of the problem greatly.\\

The matrices $\hat{\cE}_{j,\gamma_0}^{Q}$ and $\hat{\cE}_{j,\gamma_0}^{D}$ can be decomposed as follows
\be\label{simple_split}
\hat{\cE}_{j,\gamma_0}^{Q} = \half \sum_a \sum_{\alpha_j} \hat{Q}_{j,\gamma_0}(a,\alpha_j) \Sp \mbox{and} \Sp \hat{\cE}_{j,\gamma_0}^{D} = \half \sum_a \sum_{\alpha_j} \hat{D}_{j,\gamma_0}(a,\alpha_j)
\ee
where the matrices $\hat{Q}_{j,\gamma_0}(a,\alpha_j)$ and $\hat{D}_{j,\gamma_0}(a,\alpha_j)$ are now  two dimensional matrices in the space defined by $\oket{a_{\gamma_0}}$ and 
$\oket{a^{\alpha_{j}}_{\gamma_0}}$. To simplify notation we also define the vector $\oket{-^{\alpha_j}_{a_{\gamma_0}}} = \oket{a_{\gamma_0}} - \theta_{\alpha,\gamma}\oket{a^{\alpha_{j}}_{\gamma_0}}$. \\

We first consider the matrices $\hat{Q}_{j,\gamma_0}(a,\alpha_j)$: It can be verified easily that the functions $G_j(a)$ satisfy the identity $G_j(a^{\alpha_j}) = G_j(a)e^{\beta \omega^{\alpha_j}(a)}$ for all $\alpha_j$ at site $j$. This implies that $G_j(a^{\alpha_j}) \rho_a = G_j(a) \rho_{a^{\alpha_j}}$ so that $B^{Q}_{\alpha,\gamma}(a) = B^{Q}_{\alpha,\gamma}(a^\alpha)$. A simple regrouping of the sum in Eqn. (\ref{DirichletMatrix}) yields Eqn. (\ref{simple_split}) when we define

\bq\label{smallQ}
\hat{Q}_{j,\gamma}(a,\alpha) &=& \frac{1}{8}\left(G_j(a^\alpha) - G_j(a^{\alpha \gamma}) \right)^2\sqrt{\rho_a\rho_{a^\gamma}}\oproj{a_{\gamma}} \no 
					    &+& \frac{1}{8}\left(G_j(a) - G_j(a^{\gamma}) \right)^2\sqrt{\rho_{a^\alpha}\rho_{a^{\alpha\gamma}}}\oproj{a^\alpha_{\gamma}} \no 
					    &+& \frac{1}{4} G_j(a^\alpha) G_j(a^{\alpha\gamma})\sqrt{\rho_{a}\rho_{a^{\gamma}}}\oproj{-^\alpha_{a_\gamma}}.
\eq

The matrices $\hat{D}_{j,\gamma_0}(a,\alpha_j)$  for the Davies generator can be brought into a similar form: Since the rate function satisfies the KMS-condition: i.e. $h^\alpha( - \omega) = h^\alpha(\omega) e^{-\beta\omega}$, we have that $h^\alpha(\omega^\alpha(a))\rho_a = h^\alpha(\omega^\alpha(a^\alpha))\rho_{a^\alpha}$ so that  $B^{D}_{\alpha,\gamma}(a) = B^{D}_{\alpha,\gamma}(a^\alpha)$. An important difference with Eqn. (\ref{smallQ}) is that we take an average over $\alpha_{j} = 0\dots 3$ for all diagonals and distribute these weights equally among the $\hat{D}_{j,\gamma_0}(a,\alpha_j)$. We will explain why this is necessary later. This leads to matrices 
\bq\label{smallD}
\hat{D}_{j,\gamma}(a,\alpha) 
&=& \frac{1}{8} \sum_{\alpha = 0}^3 \left(h^\alpha(\omega^{\alpha}(a))+h^\alpha(\omega^{\alpha}(a^\gamma)) \right)(1 - \delta[\omega^{\alpha}(a) - \omega^{\alpha}(a^\gamma) ]) \sqrt{\rho_a\rho_{a^\gamma}}\oproj{a_{\gamma}} \no 
&=& \frac{1}{8} \sum_{\alpha = 0}^3 \left(h^\alpha(\omega^{\alpha}(a^\alpha))+h^\alpha(\omega^{\alpha}(a^{\alpha\gamma})) \right)(1 - \delta[\omega^{\alpha}(a^\alpha) - \omega^{\alpha}(a^{\alpha \gamma})]) \sqrt{\rho_{a^\alpha}\rho_{a^{\alpha\gamma}}}\oproj{a^\alpha_{\gamma}} \no 
&+& \Sp h^\alpha(\omega^\alpha(a)) \delta[\omega^{\alpha}(a) - \omega^{\alpha}(a^\gamma) ] \sqrt{\rho_{a}\rho_{a^{\gamma}}}\oproj{-^\alpha_{a_\gamma}}.
\eq
Note, that the $\delta$ function still appears in this definition. It is therefore important, that we perform a case analysis when comparing the matrices.\\

We are now finally in the position to state a simple comparison between these two matrices. We find that we can always bound 
\be\label{minimal_supp}
	 \max_{a}  2 \frac{G^2_j(a)}{h_{\min}} \hat{D}_{j,\gamma}(a,\alpha) \geq \hat{Q}_{j,\gamma}(a,\alpha),
\ee
for all $\gamma,\alpha,j$ and $a$, so that lifting this constant out of the sum, implies the theorem. To this end we need to consider three different cases. We group them according to 
the values of $\omega^\alpha(a)$.
\begin{enumerate}

\item 
The first case we consider is when $\omega^\alpha(a) = \omega^\alpha(a^\gamma)$ for all $\alpha = 0\ldots 3$. In this case we have that $ \delta[\omega^{\alpha}(a) - \omega^{\alpha}(a^\gamma) ] =1$  as well as $G_j(a^\alpha) = G_j(a^{\alpha \gamma})$ for all $\alpha$. Therefore, $\hat{D}_{j,\gamma}(a,\alpha)$ and $\hat{Q}_{j,\gamma}(a,\alpha)$ simplify greatly since the purely diagonal contributions vanish and we have that
\be
\hat{D}_{j,\gamma}(a,\alpha) = 	h^\alpha(\omega^\alpha(a)) \sqrt{\rho_{a}\rho_{a^{\gamma}}}\oproj{-^\alpha_{a_\gamma}} \Sp \mbox{and} \Sp \hat{Q}_{j,\gamma}(a,\alpha) = \frac{1}{4} G_j(a^\alpha) G_j(a^{\alpha\gamma})\sqrt{\rho_{a}\rho_{a^{\gamma}}}\oproj{-^\alpha_{a_\gamma}}.
\ee
Since the resulting matrices are up to proportionality the same projectors, we only need to compare the coefficients to obtain the estimate $\tau_j \geq \max_{a,\alpha} \frac{G_j(a^\alpha) G_j(a^{\alpha\gamma})}{4 h^\alpha(\omega^\alpha(a))} \geq  \max_{a}  2 \frac{G^2_j(a)}{h_{\min}}$. The last inequality is obtained by minimizing the numerator and the denominator independently. 

\item 
Now let us assume that all $\omega^\alpha(a) \neq \omega^\alpha(a^\gamma)$ so that none coincide. From this we have that  $ \delta[\omega^{\alpha}(a) - \omega^{\alpha}(a^\gamma) ] =0$ and $\hat{D}_{j,\gamma}(a,\alpha)$ is purely diagonal. Note that the diagonals can be bounded by $ \frac{1}{8} \sum_{\alpha = 0}^3 \left(h^\alpha(\omega^{\alpha}(a))+h^\alpha(\omega^{\alpha}(a^\gamma)) \right)\sqrt{\rho_a\rho_{a^\gamma}} \geq h_{\min}\sqrt{\rho_a\rho_{a^\gamma}}$ and $\frac{1}{8}\left(G_j(a^\alpha) - G_j(a^{\alpha \gamma}) \right)^2\sqrt{\rho_a\rho_{a^\gamma}} \leq \frac{1}{8}\max_{j,a} G^2_j(a)\sqrt{\rho_a\rho_{a^\gamma}}$, so that
\bq
&&\hat{D}_{j,\gamma}(a,\alpha) \geq h_{\min} \sqrt{\rho_a\rho_{a^\gamma}} \1_{a,a^\alpha} \geq  \half h_{\min} \sqrt{\rho_a\rho_{a^\gamma}}\left(\1_{a,a^\alpha} + \half \oproj{-^\alpha_{a_\gamma}}\right) \;\; \mbox{and} \no
&&\hat{Q}_{j,\gamma}(a,\alpha) \leq \frac{1}{8}\max_{j,a} G^2_j(a)\sqrt{\rho_a\rho_{a^\gamma}}\1_{a,a^\alpha} + \frac{1}{4} G_j(a^\alpha) G_j(a^{\alpha\gamma})\sqrt{\rho_{a}\rho_{a^{\gamma}}}\oproj{-^\alpha_{a_\gamma}}
\eq
as semidefinite inequalities. Since we have brought the matrices in a similar form again, we only need to read off the coefficients to obtain the two estimates $\tau_j \geq \max_{a}  \frac{2 G^2_j(a)}{8 h_{\min}}$ for the first term and $\tau_j \geq \max_{a}  \frac{G^2_j(a)}{h_{\min}}$ for the second. Both bounds are again consistent with $\tau_j \geq  \max_{a}  2 \frac{G^2_j(a)}{h_{\min}}$

\item
The final case to consider is when the Bohr frequencies $\omega^\alpha(a),\omega^\alpha(a^\gamma)$ coincide for some $\alpha$ while for others they do not. For the $\alpha,a$ where the Bohr frequencies do not coincide, the analysis is similar to  case $2$ studied above (up to a factor of 2 which yields the lower bound). For the $\alpha,a$  with  $\omega^\alpha(a) = \omega^\alpha(a^\gamma)$ we have a different splitting, where
\bq
&&\hat{D}_{j,\gamma}(a,\alpha) \geq \frac{1}{4} h_{\min} \sqrt{\rho_a\rho_{a^\gamma}} \1_{a,a^\alpha} + \frac{1}{2}h_{\min}\sqrt{\rho_{a}\rho_{a^{\gamma}}} \oproj{-^\alpha_{a_\gamma}}\;\; \mbox{and} \no
&&\hat{Q}_{j,\gamma}(a,\alpha) \leq \frac{1}{8}\max_{j,a} G^2_j(a)\sqrt{\rho_a\rho_{a^\gamma}}\1_{a,a^\alpha} + \frac{1}{4} G_j(a^\alpha) G_j(a^{\alpha\gamma})\sqrt{\rho_{a}\rho_{a^{\gamma}}}\oproj{-^\alpha_{a_\gamma}}
\eq
It now becomes clear why the redistribution and averaging over $\alpha$ in the definition of the diagonals of $\hat{D}_{j,\gamma}(a,\alpha)$ has become necessary. Had we not performed the sum, the diagonal contribution in $\hat{D}_{j,\gamma}(a,\alpha)$ would have disappeared and the range of  $\hat{Q}_{j,\gamma}(a,\alpha)$  could not be supported by $\hat{D}_{j,\gamma}(a,\alpha)$ making it impossible to find a finite support number.  However, with this splitting we also find that $\tau_j \geq  \max_{a}  2 \frac{G^2_j(a)}{h_{\min}}$ suffices.
\end{enumerate}

From the analysis of these three different cases we have that the bound in Eqn. (\ref{minimal_supp}) holds. If we now define $\tau = \max_{j;a} 2 \frac{G^2_j(a)}{h_{\min}}$, we have that
for all cosets $[\gamma_0]$ and all $j$ in the decomposition, that $\tau \hat{\cE}_{j,\gamma_0}^{D} \geq \hat{\cE}_{j,\gamma_0}^{Q}$ which then translates to the final bound in the lemma.
\qed}


\section{Bound on $\kappa$}

We now proceed to estimate the $a^j_{kl}$ for the heat-bath map under the assumption that the Hamiltonian is of commuting Pauli form. Before we estimate the constant $\kappa$, we state bounds on the $a^{j}_{kl}$ in terms of suitable norms of $\gamma_j$ and $\partial_k \gamma_j$ which will then be bounded in a later step.

We denote by $N_j$ the set of qubits that share a stabilizer generator $g_k$ with site $j$. That is $l \in N_j$ if for some $g_k$ both $l,j \in \mbox{supp}(g_k)$. Moreover, we denote by $B_l$ the ball that is generated by a chosen order. That is, we order all sites $l \in N_j$, with the constraint that $j$ itself is the smallest element. Then $B_l$ denotes all elements that are smaller or equal to $l$ with respect to this ordering. 

\begin{lemma}\label{bound-on-a} Let $\cQ$ denote the heat Bath generator of Eqn. (\ref{eqn:heatbath}), then we have: 
\be
	 ||[\partial_k,\cL_j](f)|| \leq \sum_l a^j_{kl} ||\partial_l(f)||,
\ee
where 
\bq
a^j_{kl} = \left\{ \begin{array}{ll} 4 \|\gamma_j \| \sum_{s \in B(j)^c_l} \|\partial_s \gamma_j \| &: \Sp \mbox{if} \Sp k \in N_j \vspace{0.2cm} \\  0  &: \Sp \mbox{otherwise}. \end{array} \right.
\eq
Note that $B(j)_l$ refers to the labels $l$ that are in the chosen ordering in $B(j)_l \subset N_j$ smaller than $s < l$. The sum is taken over the complement $B(j)^c_l = \Lambda \backslash B(j)_l$.
\end{lemma}

\vspace{0.3cm}
\proof{  {It} follows immediately from the definition of $\cQ_j = \bE_j[f] - g$ in terms of the non-commutative expectation value that $[\partial_k,\cL_j](f) = 0$ whenever $k \notin N_j$. 
This directly implies the second line in the definition of the  $a^j_{kl}$. To obtain the bound for the case where $k \in N_j$ we proceed by recalling the definition of the heat-bath generator 
\be
 \cQ ( f) \assign \frac{1}{4}\sum_{j, \alpha} \left(A_j^{\alpha \dag } f A_j^{\alpha} - \frac{1}{2} \{ f, A_j^{\alpha \dag} A_j^{\alpha}\}\right),
\ee
where each of the Lindblad operator is defined as $A_j^{\alpha} = \gamma_j \sigma^\alpha$ with $\gamma_j = \sum_a G(a)_j P(a)$.  {We only need to consider the local summands in the expressions for $a^j_{kl}$.} Every Lindblad generator can  conveniently be written as the sum of two commutators
\be
\cQ_j = \frac{1}{4} \sum_{\alpha}\cL_j^\alpha(f) =  \frac{1}{4} \sum_{\alpha} \left(\half [A_j^{\alpha \dag }, f]A_j^{\alpha} + \half A_j^{\alpha \dag } [f, A_j^{\alpha}] \right).
\ee
From this we can directly evaluate the commutator for every summand $\cL_j^\alpha$ 
\bq\label{evalCommutator}
[\partial_k,\cL^\alpha_j](f) =  
  \half \left( [A_j^{\alpha \dag }, \ptr{k}{f}]A_j^{\alpha} + A_j^{\alpha \dag } [ \ptr{k}{f}, A_j^{\alpha}] \right)
 -\half \left( \ptr{k}{[A_j^{\alpha \dag }, f]A_j^{\alpha}} + \ptr{k}{A_j^{\alpha \dag } [ f, A_j^{\alpha}]} \right) 	
\eq
 {We now proceed by applying a trick in Ref. \cite{ZiggyFirst} to evaluate the commutator.} Note that $[g -\tr{g},f - \tr{f}] = [g,f]$ for any observables $f,g$.  {Then, }we can choose an ordering of lattice sites starting at site $i$, i.e. $i$ is the smallest element in $\Gamma_i$, to reach the following operator identities,
\bq
g - \tr{g}  =  \sum_{m=1}^N \ptr{[l_1,\ldots,l_m]^c}{\partial_{l_m}g} \Sp \mbox{and} \Sp
f  - \tr{f}   =  \sum_{n=0}^{N-1}  \ptr{[l_1,\ldots,l_n]}{\partial_{l_{n+1}}f}, 
\eq
where we denote the partial trace over the sites $\{l_1,\ldots,l_n\}$ by $\ptr{[l_1,\ldots,l_n]}{\cdot}$ and the complement by $[l_1,\ldots,l_m]^c = \Lambda \backslash \{l_1,\ldots,l_m\}$.
Note that the shelling of the space works in the opposite direction for $f$ and $g$ so that
\bq\label{commutatorID}
\left[g,f\right] &=& \sum_{m=1}^N \sum_{n=0}^{N-1} \left[\ptr{[l_1,\ldots,l_m]^c}{\partial_{l_m}g}, \ptr{[l_1,\ldots,l_n]}{\partial_{l_{n+1}}f}\right] \no
		    &=& \sum_{n=0}^{N-1}  \sum_{m \geq n + 1}^N \left[\ptr{[l_1,\ldots,l_m]^c}{\partial_{l_m}g}, \ptr{[l_1,\ldots,l_{n-1}]}{\partial_{l_{n+1}}f}\right].\no
\eq
In the last line we used the fact that partial trace yields an identity on the subsystem so  the commutator $\left[\ptr{[l_1,\ldots,l_m]^c}{\partial_{l_m}g}, \ptr{[l_1,\ldots,l_{n-1}]}{\partial_{l_{n+1}}f}\right] = 0$ for all $m < n+1$. This commutator indetity can now be used to bring the terms in Eqn. (\ref{evalCommutator}) into a suitable form and apply  norm bounds  {from there}.  We apply Eqn. (\ref{commutatorID}) to Eqn. (\ref{evalCommutator}) and obtain
\bq
[\partial_k,\cL^\alpha_j](f) =  \sum_{n=0}^{N-1}  \sum_{m \geq n + 1}^N 
&&   \half {\Big (} \left[\ptr{[l_1,\ldots,l_m]^c}{\partial_{l_m}A_j^{\alpha \dag}}, \ptr{[l_1,\ldots,l_n],k}{\partial_{l_{n+1}}f}\right]A_j^{\alpha} \no 
&&  + A_j^{\alpha \dag } \left[\ptr{[l_1,\ldots,l_n],k}{\partial_{l_{n+1}}f}, \ptr{[l_1,\ldots,l_m]^c}{\partial_{l_m}A_j^{\alpha}}\right] \Big{)} \no
&&   -\Big{(} \ptr{k}{\left[\ptr{[l_1,\ldots,l_m]^c}{\partial_{l_m}A_j^{\alpha \dag}}, \ptr{[l_1,\ldots,l_n]}{\partial_{l_{n+1}}f}\right]A_j^{\alpha}} \no
&&  + \ptr{k}{A_j^{\alpha \dag}  \left[ \ptr{[l_1,\ldots,l_n]}{\partial_{l_{n+1}}f},\ptr{[l_1,\ldots,l_m]^c}{\partial_{l_m} A_j^{\alpha}}\right]} \Big{)}. 
\eq
The operator norm for two matrices is both sub multiplicative, ($\|g f\| \leq \|g\|\|f\|$), and contractive under the partial trace ($\|\ptr{S}{g}\| \leq \|g\|$). Furthermore the triangle
inequality can be used to bound $\|[f,g]\| \leq 2\|fg\|$.  {Hence, for $l \neq j$ }
\be
	\|\partial_l A_j^{\alpha \dag}\| = \|\partial_l A_j^{\alpha}\| \leq \|\partial_l \gamma_j\|,
\ee
since $\partial_l(\gamma_j \sigma^\alpha_j) = (\partial_l\gamma_j)\sigma_j^\alpha$ because Pauli operators acting on different sites commute. Moreover we have that $\|\sigma_j^\alpha\| =1$. The norm bound for $\partial_j A_j^{\alpha}$ is not so direct, since the action of the Pauli matrix $\sigma^\alpha_j$ and $\partial_j$ do not commute any longer. However, we observe that this occurs only for the first summands in $m$ which are of the form $\ptr{[j]^c}{\partial_{j}A_j^{\alpha}} = \ptr{[j]^c}{A_j^{\alpha}} - \tr{A_j^\alpha}$. Since this term appears in the commutator and $\tr{A_j^\alpha} \propto \1$, we have that this term vanishes in the commutator. Therefore we 
can immediately state the bound
\be
\|[\partial_k,\cL^\alpha_j](f)\| \leq \sum_{n=0}^{N-1}  \sum_{m \geq n + 1}^N 4 \|\gamma_j\| \|\partial_{l_m} \gamma_j\| \|\partial_{l_{n+1}}f\|
\ee 
Now recall that $\|[\partial_k,\cL_j](f)\| \leq \frac{1}{4}\sum_{\alpha} \|[\partial_k,\cL^\alpha_j](f)\| $ due to the triangle inequality. Relabeling the summands $s = l_{m}$ and $l = l_{n+1}$, 
we are left with the estimate 
\be
	\|[\partial_k,\cQ_j](f)\| \leq \sum_{l=1}^N \underbrace{\left( \sum_{s \geq l}^N 4 \|\gamma_j\| \|\partial_{s} \gamma_j\|\right)}_{a_{kl}^j} \|\partial_{l}f\|,
\ee
which  {yields the desired bound.}  \qed}

\vspace{0.3cm}
Since we now have an estimate for the  {$a_{kl}^l$ - constants in terms of the $\gamma_j$ matrices}, we only need to evaluate the operator norms of $\gamma_j$ and $\partial_m \gamma_j$. In principle the sum in the definition of the $a^j_{kl}$ has to be  taken over the full lattice with the exception of a ball $B_l$ that surrounds the site $j$. However, it will become evident that the summands $a^j_{kl}$ will 
vanish, whenever  $l \notin N_j$ any longer. This will follow from the property that $\|\partial_m \gamma_j\| = 0$, once $m \notin N_j$, since the sum is empty  {when} the ball $B_l$ is larger than $N_j$.

\begin{proposition}\label{kappaBnd}
The condition number $\kappa$ is bounded by
\be
	\kappa(\beta) = \max_j \frac{3\left(|N_j| -1\right)}{\frac{1}{4} \sum_{\alpha_j=0}^3 e^{-2 \beta|e(\alpha_j)|}} \sum_{l \in N_j} \sum_{m > l} \left(e^{2\beta\epsilon(m,j)} -1\right),
\ee
where 
\be
	\epsilon(m,j) = \max_{\alpha_j,\tau_m}\sum_k J_k e_k(\alpha_j)e_k(\tau_m).
\ee
\end{proposition}

{\it Remark:} Before we proceed with the proof, let us note that  {a simpler expression can be obtained}, which does not depend on the syndromes $e(\alpha)$. The only relevant model dependent parameters are the number of qubits $|N_j|$ that are connected to the qubit $j$ through adjacent stabilizers, and the number of stabilizers that are shared 
between adjacent qubits $m,j$ denoted by $|S_j \cap S_m|$. We have that, 
\be
	\kappa(\beta) = \max_j 3(|N_j| - 1) e^{2 \beta J |S_j|} \sum_{l \in N_j} \sum_{k > l}(e^{2 \beta J| S_j \cap S_k |} -1)
\ee
which follows from the upper bound on $\epsilon(m,j)$ in terms of the shared number of stabilizers. Note that we also gave a bound on  ${\frac{1}{4} \sum_{\alpha_j=0}^3 e^{-2 \beta|e(\alpha_j)|}}$ by the smallest summand. The smallest summand can simply be bounded by the exponential of the total number of stabilizers $|S_j|$ that are supported on site $j$.\\

\proof{ The central ingredient for obtaining bounds on the $a^j_{kl}$ stems from finding suitable norm bounds on the $\gamma_j$ and $\partial_m \gamma_j$. Note that the operator norm
of the matrix $\gamma_j$ is evaluated immediately, since the projectors $P(a)$ are mutually orthogonal. so that 
\be\label{nrm-bnd}
	\|\gamma_j\| = \max_a G_j(a)  = \left( \frac{1}{4} \sum_{\alpha_j} e^{-2\beta J|e(\alpha_j)|} \right)^{-1/2}.
\ee
Moreover, we can estimate the norm of $\partial_l \gamma_j$ by the following sequence of steps
\bq
\|\partial_m \gamma_j\| = \max_a \left | \frac{1}{4} \sum_{\tau_m = 0}^3 (G_j(a) - G_j(a^\tau)) \right| \leq \frac{3}{4} \max_{a,\tau_m} \left | G_j(a) - G_j(a^{\tau_m})  \right|
\eq
where we have made use of the fact that we can bound every summand by the maximal except for $\tau_m = 0$, which always vanishes. Now it is easy to see that 
\be
	\max_{a,\tau_m} \left | G_j(a) - G_j(a^{\tau_m})  \right| \leq \|\gamma_j\| \max_{a,\tau_m} \left | \frac{G_j(a)}{G_j(a^{\tau_m})} -1  \right|
\ee
We now assume that $\frac{G_j(a)}{G_j(a^{\tau_m})}\geq 1$ so we can drop the absolute values. If this is not the case we can map $a \raw a^\tau$ and flip the sign so that the assumption holds again. This is possible since we are considering the maximum over all $a$. So the two cases are in fact symmetric. Bounding both terms for every summand in $\cG$ we can see that 
we always have
\be
	 \max_{a,\tau_m} \frac{G_j(a)}{G_j(a^{\tau_m})} \leq \max_{\tau_m, \alpha_j} e^{2\beta \sum_k J_k e_k(\alpha_j)e_k(\tau_m)}. 
\ee
since the exponential is a monotone function the bound is nothing but $\exp(2\beta\epsilon(m,j))$. Hence we obtain the upper bound to the norm
\be
	\|\partial_m \gamma_j\| \leq \frac{3}{4}\|\gamma_j\|\left(e^{2\beta\epsilon(m,j)} - 1\right)
\ee
from this bound it can be seen easily that $\|\partial_l\gamma_j\| = 0$, whenever $l \notin N_j$ as would be expected. This holds, since $\epsilon(l,j) = 0$, when 
the sites $l,j$ do not share a stabilizer.  Hence using the formula from lemma \ref{bound-on-a} for the $a^{j}_{kl}$  we have
\bq
a^{j}_{kl} \leq \left\{ \begin{array}{ll} 3\|\gamma_j\|^2 \sum_{m \in N_j\backslash B_l} \left(e^{2\beta\epsilon(m,j)} - 1\right) &: \Sp \mbox{if} \Sp k \in N_j \vspace{0.2cm} \\  0  &: \Sp \mbox{otherwise}. \end{array} \right.
\eq
Now let us recall that the constant $\kappa(\beta)$ is determined in therms of the sum
\be
\kappa(\beta) \leq \max_j \sum_{k \in j^c} \sum_l a_{kl}^j = \max_j \left( |N_j| - 1\right) 
\sum_{l \in N_j} 3\|\gamma_j\|^2 \sum_{m \in N_j\backslash B_l} \left(e^{2\beta\epsilon(m,j)} - 1\right).
\ee
 {The} sum in $k$ vanishes for summands $k\notin N_j$ as already observed in  lemma \ref{bound-on-a}. Since the summands do not  depend explicitly on $k$, we just estimate this sum in terms of $|N_j| - 1$ because $j$, the center, is excluded. Moreover, we have observed that $\partial_m \gamma_j =0$, whenever $m \notin N_j$. This leads to the 
fact the $a^j_{kl}$ vanish for all $l \notin N_j$ because the  {set} $N_j \backslash B_l$ is empty. This justifies the restriction of the summands in the second equality. Now recall that the norm $\|\gamma_j\|$ is given by Eqn. (\ref{nrm-bnd}) so that we are left with the expression as stated in the lemma.  \qed}


\section{Gap from ergodicity}
In this section we show that strong ergodicity of the semigroup (Eqn. (\ref{ergodicity})) implies that the generator is gapped. Indeed, 

\begin{lemma}\label{Gap-erg}
Let $\cL$ be the generator of a primitive reversible quantum dynamical semigroup. If there exist constants $c,\kappa$ such that 
\be || e^{t \cL}(f)-\tr{\rho f}||_\infty \leq c ||| f ||| e^{-t (1-\kappa)},\ee
for all observables $f$, then the generator has a gap which is lower bounded by $1-\kappa$. 
\end{lemma}
\proof{
The proof hinges on the use of interpolation theorems for $\bL_p\rightarrow\bL_q$ norms. A non-commutative $\cL_p$ norm is defined with respect to a full rank density matrix $\rho>0$ as $||f||^p_{p,\rho}=\tr{|\rho^{1/2p}f\rho^{1/2p}|^p}$ (see Refs. \cite{GibbsSamplers,ZiggyFirst,LogSobolev} for more details). The relevant inner product is $\avr{f,g}_\rho=\tr{\rho^{1/2}f^\dag \rho^{1/2}g}$. 
We will first show that 
\be \|e^{t \cL}-\bE^\infty\|_{1-1,\rho}=\|e^{t \cL}-\bE^\infty\|_{\infty - \infty,\rho},\ee
where $\bE^\infty\equiv \lim_{t\rightarrow\infty}e^{t\cL}$. 
Indeed, from the variational characterization of norms, and reversibility of the semigroup, 

\begin{align}  
\|e^{t \cL}-\bE^\infty\|_{1-1,\rho}=& \sup_{||f||_{1,\rho}\leq 1}||(e^{t \cL}-\bE^\infty)(f)||_{1,\rho} \nonumber\\
=& \sup_{||g||_\infty\leq 1,||f||_{1,\rho}\leq 1} \avr{g,(e^{t \cL}-\bE^\infty)(f)}_\rho \nonumber\\
=& \sup_{||g||_\infty\leq 1,||f||_{1,\rho}\leq 1}\avr{f,(e^{t \cL}-\bE^\infty)(g)}_\rho\nonumber\\
=& \sup_{ ||g||_\infty\leq 1}||(e^{t \cL}-\bE^\infty)(g)||_\infty\nonumber\\
=& \|e^{t \cL}-\bE^\infty\|_{\infty - \infty}\nonumber,
\end{align}
where we have used that $\bE^{\infty}$ and $e^{t\cL}$ are reversible with respect to the inner product $\avr{f,g}_\rho=\tr{\rho^{\half}f^\dag\rho^{\half}g}$. 
Now, invoking an extension of the Riez-Thorin theorem \cite{RiezThorin} (see Lemma \ref{RTinterpol}), we get for any linear operator $T$,
\be ||T(f)||_{2,\rho}\leq 2 \sqrt{2N_1N_\infty} ||f||_{2,\rho}, \ee 
whenever $\|T\|_{1-1,\rho}\leq N_1$ and $ \|T\|_{\infty - \infty}\leq N_\infty$. 

Therefore, setting $T:=e^{t\cL}-\bE^\infty$, and observing that
\be || e^{t\cL}(f)-\tr{\rho f}||_\infty \leq c |||f|||e^{-(1-\kappa) t}\leq 2c |\Lambda_f|~||f||_\infty e^{-(1-\kappa) t},\ee
where $|\Lambda_f|$ is the support of observable $f$ and
\be || e^{t\cL}(f)-\tr{\rho f}||_{1,\rho} \leq c |\Lambda_f|~||f||_{1,\rho} e^{-(1-\kappa) t},\ee
we get
\be || e^{t\cL}(f)-\tr{\rho f}||_{2,\rho} \leq 2c |\Lambda_f|~||f||_{2,\rho} e^{-(1-\kappa) t}.\ee

Then, by picking the operator $f_\lambda$ such that $f_\lambda-\tr{\rho f_\lambda}$ is the eigenvector corresponding to the spectral gap $\lambda$, we get
\bea || e^{t\cL}(f_\lambda)-\tr{\rho f_\lambda}||_{2,{\rho}} &=& || f_\lambda-\tr{\rho f_\lambda}||_{2,\rho}e^{-t \lambda}\\
&\leq& c' |\Lambda_f|~||f_\lambda||_{2,\rho} e^{-(1-\kappa) t}\eea
This inequality must holds for all $t>0$, hence $\lambda\geq 1-\kappa$. \qed
}

\begin{lemma}[Ref. \cite{RiezThorin}]\label{RTinterpol}
Let $T$ be a bounded linear operator, and assume that there exist constants $N_p,N_q$ such that for $p\leq q$, and any (hermitian) operator $f$ we have 
\be ||T(f)||_{p,\rho}\leq N_p ||f||_{p,\rho} ~~~~~~~~~~~~ and ~~~~~~~~~~~~~~  ||T(f)||_{q,\rho}\leq N_q ||f||_{q,\rho}\ee

then for any $p\leq r\leq q$ we get 
\be ||T(f)||_{r,\rho}\leq \gamma N^\delta_pN^{1-\delta}_q ||f||_{r,\rho}, \ee
with 
\be \delta = \frac{p(q-r)}{r(q-p)} ~~~~~~~~~~~~ and ~~~~~~~~~~~~~~ \gamma= 2\left( \frac{r(q-p)}{(r-p)(q-r)}\right)^{1/r}\ee
\end{lemma}

\end{appendix}
\end{widetext}
\end{document}